\documentclass[10pt, conference, compsocconf]{IEEEtran}
\IEEEoverridecommandlockouts                              
\usepackage{graphics,color} 
\usepackage{subfigure}
\usepackage{rotating}
\usepackage{algorithmic}
\usepackage{comment}
\usepackage{algorithm,amssymb}
\usepackage{epsfig}
\usepackage[cmex10]{amsmath}
\usepackage{amsthm}
\usepackage{bm}
\usepackage{cite}
\usepackage{textcomp}

\usepackage{hyperref}
\hypersetup{
  hidelinks, 
}

\newcommand {\eeq} {\end{equation}}
\newcommand {\barr} {\begin{array}}
\newcommand {\earr} {\end{array}}
\newcommand {\bear} {\begin{eqnarray}}

\newcommand {\namep} {Foggy\relax}
\begin{document}
\title{\namep: a platform for workload orchestration in a Fog Computing environment\footnote{}}

\author{Daniele Santoro, Daniel Zozin, Daniele Pizzolli, Francesco De
  Pellegrini, Silvio Cretti \\ FBK CREATE-NET, via alla Cascata 56/D, 38123
  Trento, Italy. Contact: dsantoro@fbk.eu}

\maketitle

\begin{abstract}
\label{sec:orge8f33df}
In this paper we present Foggy, an architectural framework and software
platform based on Open Source technologies. Foggy orchestrates application
workload, negotiates resources and supports IoT operations for multi-tier,
distributed, heterogeneous and decentralized Cloud Computing systems.

Foggy is tailored for emerging domains such as 5G Networks and IoT, which
demand resources and services to be distributed and located close to data
sources and users following the Fog Computing paradigm. Foggy provides a
platform for infrastructure owners and tenants (i.e., application providers)
offering functionality of negotiation, scheduling and workload placement
taking into account traditional requirements (e.g. based on RAM, CPU, disk)
and non-traditional ones (e.g. based on networking) as well as diversified
constraints on location and access rights. Economics and pricing of resources
can also be considered by the Foggy model in a near future.

The ability of Foggy to find a trade-off between infrastructure owners' and
tenants' needs, in terms of efficient and optimized use of the infrastructure
while satisfying the application requirements, is demonstrated through three
use cases in the video surveillance and vehicle tracking contexts.
\end{abstract}


\section{Introduction}
\label{sec:intro}
In the last few years, Cloud computing has undergone a deep transformation
driven by the technological evolution, namely new containerization techniques,
and by new requirements imposed by emerging 5G and IoT domains. From the
technological standpoint, the operating system virtualization, i.e. the
containerization complements the traditional virtual machines one, because it
is more lightweight, flexible, and portable. Thus, developers and cloud
providers can devise innovative architectural patterns such as microservices
and novel paradigm like Infrastructure as Code. At the same time, emerging 5G
networks and pervasive IoT technologies demand for distributed and
decentralized support by Cloud services: Cloud Computing must escape from
centralized data centers and Cloud services should be provided closer to the
users or data sources. This corresponds to the vision of Fog Computing
\cite{bonomi}.

Several advantages of such an approach, including real time responses, low
network latency and bandwidth usage, fault tolerance and support to data
privacy, are well known in literature \cite{cloudlet}. Scenarios in which
these new cloud technologies can be applied include IoT, Automotive, Industry
4.0, Tactile Internet applications. At the same time, a whole new set of
stakeholders can benefit, including, e.g., cloud providers, telecommunication
providers, IoT providers, cloud integrators, innovative industries and
application providers.

Our reference application scenario is a smart city one, where a Fog
infrastructure is connected with IoT sensors and cameras. The infrastructure
belongs to a given owner (O) that offers resources and services on this
infrastructure to different tenants (T), wanting to deploy innovative
applications and competing for the same resources.  Tenants can negotiate
resources and services which the infrastructure owners can guarantee and
reserve to specific applications. In this scenario, business-related problems
that should be addressed include: i) maximize the usage of the infrastructure
avoiding over provisioning (O); ii) satisfy the contractual SLA, i.e. minimize
SLA violations across tenants (O); iii) maximize revenues (O); iv) negotiate
requested resources, services and corresponding SLA (T); v) adapt to changing
requirements in a flexible manner (e.g. burst of data, need for real-time
processing, more workload to process, etc.) (T); vi) minimize economic costs
(T).

From a technological point of view, conversely, handling deployment and
operations and managing workload orchestration and efficient placement in a
distributed, heterogeneous, decentralized, multi-tier cloud environment adds
many “degrees of freedom” to similar problems with respect to a centralized
and homogeneous cluster of resources \cite{mobility_aware_scheduling}. In
fact, devices and services are scattered on the territory, and not always
connected to data centers (e.g., a public cloud) through reliable and
homogeneous network connections. In this context, the process of scheduling
applications should account for a wider range of different parameters. It is
reasonable to foresee a whole new class of policies with respect to the legacy
ones currently used to allocate computing and network resources in traditional
systems.  Such policies should indeed cover jointly context-awareness,
location detection, and network performance.

Even though many scientific papers, envisioning a tight coupling and
integration of IoT, Cloud and Future Network technologies, have been published
in recent years \cite{botta2016integration}, very few effort has been carried
out on the development of platforms to make this integration possible and
optimized. Foggy, as an evolution of a previous work \cite{cloud4iot}, is
meant to improve the aforementioned integration and the experimentation on
workload management, thanks to a model of ICT resources that considers not
only computation and storage ones, but also the kind, location, spatial
distribution and the networking among them, leveraging osmotic computing
concepts \cite{osmotic_flow}.

The architecture and implementation of Foggy will be presented in
Sec.~\ref{sec:arch} and \ref{sec:impl}. Three use cases that are meant to
elucidate the Foggy's capabilities to orchestrate workloads in a distributed
environment are then described in Sec.~\ref{sec:usecases}. Finally concluding
Sec.~\ref{sec:concl} comments on achieved results and suggests next research
directions.

\section{Architecture}
\label{sec:arch}
\begin{figure}[t]
  \centering \includegraphics[width=\linewidth]{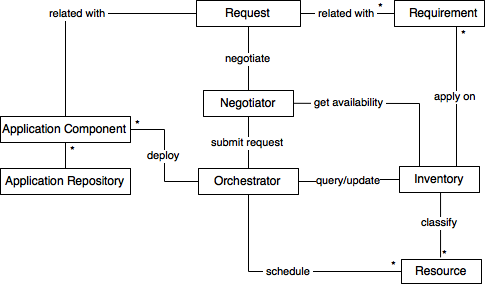}
  \vspace*{-7mm}
  \caption{Logical diagram representing the
    \namep~platform.}\label{fig:diagram}
  \vspace*{-5mm}
\end{figure}
Foggy aims at managing the workload placement in a Fog infrastructure
satisfying requirements of deployed applications that in turn request to
access and use resources and services offered by that infrastructure.

\noindent The design of Foggy platform focuses on a multi-tier Cloud Computing
context generally composed by more than three (3+) tiers specialized for Fog
environments: i) \emph{Cloud} tier which offers high resource capacity and is
generally far from the source of data; ii) \emph{Edge} \emph{Cloudlets} tier
composed by physical or virtual nodes which have medium resource capacity and
are closer to data sources; iii) \emph{Edge Gateways} tier that is composed by
nodes with low storage and computational capacity which are located very close
to data producers; iv) \emph{Swarm of Things} tier is where IoT devices
(sensors, actuators, cameras, smartphones) are hosted and where raw data is
produced. Foggy aims at controlling the first three tiers.

Fig.~\ref{fig:diagram} shows the main components of Foggy and their
relationships in the form of a logical model. In order to build this model we
considered two reference schedulers, namely, OpenStack
\cite{openstack_scheduler} and Kubernetes\cite{kubernetes_scheduler}.

\noindent A deployment \emph{Request} is of the form:
$Request=\{app\_component, [Req1, Req2, \ldots{}, ReqN]\}$ and consists of: i)
the \emph{Application component} to be deployed and ii) a set of optional
deployment \emph{Requirements}.

\noindent\emph{Application Component} is an independently deployable,
replaceable and upgradable unit of software, such as a microservice, which
plays a specific role as part of a larger application. It is generally
distributed via software container images (e.g, a Docker container image),
which are stored in an \emph{Application Repository} (e.g., a Docker
registry).

\noindent \emph{Requirement} offers a way for the tenant to specify
constraints imposed to the deployment/execution of the \emph{Application
  Component} in terms of \emph{Resources} requested and/or specific
application needs (e.g., location, access rights).\\ \emph{Resource} refers to
any computational, storage or network capacity provided by the nodes of the
infrastructure. Regarding computational and network resources, Foggy
identifies a set of profiles in order to simplify how the associated
requirements are expressed.  Thus, computational resources on a node such as
vCPUs, RAM and disk are characterized by the following usage profiles:
\emph{General purpose} (default profile), \emph{Compute optimized},
\emph{Memory optimized} and \emph{Storage optimized}. Network resources such
as bandwidth, latency and jitter are defined between an \emph{Application
  Component} (thus the node where it is executed) and a service endpoint
(e.g., a stream from a camera). They are classified with the following usage
profiles: \emph{Best Effort} (default profile), \emph{Interactive
  application}, \emph{Signaling and video streaming}, \emph{Interactive and
  real-time video}.\\ The \emph{Inventory} stores the status (i.e., resources
availability) of the distributed infrastructure together with the resources
location. It is populated with information from external systems like SDN
network orchestrators and/or IaaS managers (e.g., ONOS,
OpenStack). Information maintained by the Inventory are key for the Foggy
operations and must be preserved, for this reason it is based on a consistent,
distributed and highly available key value store, such as etcd
\cite{etcd}).\\ The \emph{Negotiator} handles the submitted \emph{Requests},
negotiating with the tenants the possibility to satisfy the associated
requirements. The resources' availability status is retrieved from the
\emph{Inventory}.

\noindent The \emph{Orchestrator}, in response to deployment requests, deploys
\emph{Application Components} on the node that best satisfy the requirements
imposed. It embeds a custom, shared state scheduler \cite{omega} which extends
the Kubernetes one and supports non-traditional requirements (i.e. beyond
computational and storage capacity).

Foggy receives deployment \emph{Requests} submissions from tenants through a
RESTful API and processes them one by one following a First Come - First
Served (FCFS) policy. First step is to go through a transaction mechanism
handled by the \emph{Negotiator}. In this phase the request from the tenant is
either accepted or rejected. The second step is responsibility of the
\emph{Orchestrator} that, by querying the \emph{Inventory}, applies filtering
and ranking rules to identify the best nodes to host the requested
\emph{Application Component}. It first i) filters the nodes that can satisfy
the requirements specified in the deployment request; then it ii) ranks the
remaining nodes according to a ”priority function”; iii) chooses the highest
from the rank results; and iv) deploys on that node the container image of the
accepted application component. Finally, the \emph{Orchestrator} updates the
\emph{Inventory} to reflect the global status of the resources.

\section{Implementation}
\label{sec:impl}
\begin{figure}[t]
  \centering \includegraphics[width=\linewidth]{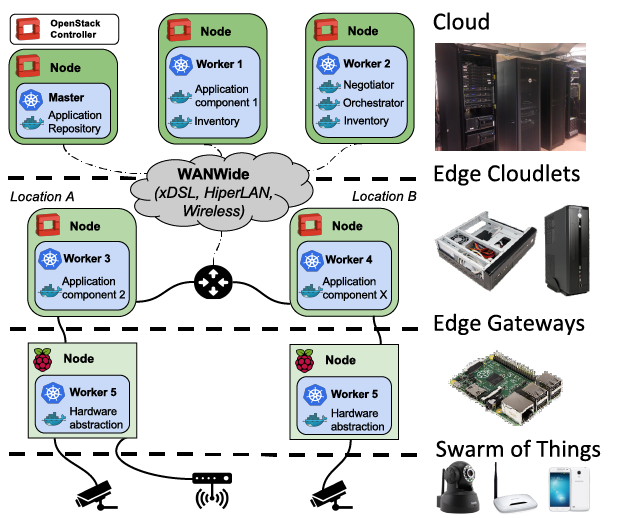}
  \vspace*{-7mm}
  \caption{Representation of the \namep~testbed: (left) deployment
    architecture in a smart city, (right) kind of hardware for each
    tier.}\label{fig:deployment}
  \vspace*{-5mm}
\end{figure}

We present hereafter the technical details of the implemented solution
including specific reference to the technologies adopted in Foggy. To support
such a flexible environment OpenStack acts as the IaaS layer, while Kubernetes
acts as the container orchestration tool. This scheme permits to achieve
application and services orchestration in a lightweight and flexible
manner. In Fig.~\ref{fig:deployment} we present the schema of our deployment:
the three main software platforms, i.e., OpenStack, Kubernetes and Foggy are
stacked one above the other.

\noindent \emph{OpenStack:} the OpenStack distributed deployment follows the
architecture proposed by the Fog/Edge Massively Distributed Cloud Working
Group \cite{femdc}.  This deployment installs OpenStack controller nodes on
the Cloud tier while compute nodes on both Cloud and Edge Cloudlets tiers,
keeping them interconnected via "WANWide" links.

\noindent \emph{Kubernetes:} a customized Kubernetes cluster is installed in a
hybrid way depending on the tier: on the Cloud and Cloudlets tiers, it is
distributed on top of OpenStack virtual machines, thus granting maximum
isolation and flexibility, while on the Edge Gateways tier it is directly
installed on the bare metal nodes due to scarce computational
resources.

\noindent \emph{Foggy:} The Foggy platform, composed by three main modules:
\emph{Negotiator}, \emph{Orchestrator} and \emph{Inventory}, is deployed by
Kubernetes using Docker containers. The \emph{Application Repository} (e.g.,
Docker registry) can be public or private and it is generally deployed on the
Cloud tier but the place where images are located can affect the start-up time
of \emph{Application components}, which in turn, are deployed by Foggy on top
of Kubernetes worker nodes.

\noindent \emph{Physical testbed:} the Cloud tier is hosted in the FBK
CREATE-NET data center, while the Edge Cloudlet tier is composed by 3 nodes
(Intel i7 CPU, 16GB RAM, 480GB SSD). At the Edge Gateways level we deployed
small and low power consumption devices (Raspberry Pi version 3) that serve as
both i) hardware abstraction layer and ii) network provider for non-IP IoT
devices. For the specific use cases demoed in this paper, the Swarm of Things
tier is composed by access points (TP-Link TL-WR740N), cameras (Tenvis
JPT3815W-HD) and smartphones (Samsung GT-I9195); indeed several other devices
could be attached to validate different scenarios.

To test Foggy, we need to control the status and performances of the
interconnections among the 3+ tiers. Using the EnOS tool \cite{femdc}, we are
able to emulate various OpenStack deployments with different kinds of
connectivity. As an example, EnOS can emulate a real world situation in which
the link between the Edge Cloudlets and the Cloud tiers is offered by an xDSL
connection with low bandwidth and high latency capabilities.

\section{Storyboard and Use Cases}
\label{sec:usecases}
As introduced earlier, the reference scenario is the smart city one where the
Fog infrastructure is owned and managed by a public entity (e.g., a
municipality). The infrastructure allows to access cameras installed on public
streets and squares. With Foggy, the municipality can lease this
infrastructure to multiple tenants. Tenants, in turn, can deploy components of
their applications close to cameras, thus leveraging the advantages of the Fog
Computing paradigm for better user experience and premier grade service.

In order to showcase the advantages of Foggy compared to an IoT platforms
based on a conventional cloud-only approach (i.e., without the Fog tiers),
three use cases have been devised.\\ In the first two, a tenant wants to
process video streams coming from cameras in order to perform face
detection. The very first represents a baseline case: a cloud-only IoT
application acquires data by streaming them directly from a camera to specific
application component running in the central cloud. The second demonstrates
how Foggy, as specified by the application deployment request, can schedule
the workload close to the data source: when face detection from a video stream
is performed close to the source, the traffic load towards the cloud is
reduced and the service level of the application is maintained even in case of
limited network performance. The third use case highlights the ability of
Foggy to orchestrate the workload based on the geographic location: the key
scenario involves privacy constraints which need to be satisfied. A tenant
wants to track vehicles location and, for privacy concerns, such sensible data
is processed only within a geographically limited area before being sent,
anonymised, to the central cloud. Deploying the tracking functionality at the
edge also increases the application resiliency by avoiding to lose data in
case of connectivity faults between Cloud and Edge.

\subsection{Cloud-only Face Detection}
\label{sec:orgcfc5d3c}
In this use-case a tenant wants to extract human faces \cite{amos2016openface}
from a video stream. The video stream is generated by a camera and forwarded
to a cloud data-center, where it is processed. No intermediate Fog tier is
used. The bandwidth has to be sufficient to stream the video produced by the
camera to the remote server in the central cloud. Scalability problems arise
when multiple cameras are used, since bandwidth requirements increase
accordingly.  The application is composed by two microservices:
$face\_detection$ and $face\_store$. The former detects and extracts faces
from a video stream while the latter stores them for persistency. This
demonstration comprises the following steps:

{\indent 1)} The tenant submits to Foggy two deployment requests:
$r_1=\{face\_detection\}$ and $r_2=\{face\_store\}$. Note that in both $r_1$
and $r_2$ no deployment requirement is specified;

{\indent 2)} Foggy, accounting for the current status of the infrastructure,
tries to deploy the microservices by satisfying the tenant requests. Since no
particular requirement is specified, both microservices are deployed on the
central cloud;

{\indent 3)} The streaming video from the camera is collected and face
detection is performed;

{\indent 4)} Detected faces are saved in an object storage.

\noindent If another application is deployed and requires video streams from
cameras, bandwidth usage increases and the performance of concurrently
deployed applications may experience degradation.

\subsection{Bandwidth-aware Face Detection}
\label{sec:bandwith_aware}
This application performs same activity as in the previous scenario. However,
now the tenant specifies the face detection microservice to be deployed close
to the data source. In this way, only the detected faces are extracted and
forwarded to the cloud, thus reducing the bandwidth consumption with respect
to the cloud-only face detection scenario. This demonstration is composed by
the following steps:

{\indent 1)} The tenant submits to Foggy two deployment requests:
$r_1=\{face\_detection, n_{svs}\}$ and $r_2=\{face\_store\}$. Where $n_{svs}$
indicates a network requirement specifying a \emph{Signaling and video
  straming} profile between the $face\_detection$ and the endpoint (e.g., the
camera).

{\indent 2)} Foggy, accounting for the current status of the infrastructure,
deploys the microservices satisfying the tenant requests. It deploys the
$face\_detection$ on a cloudlet close to the camera in order to meet the
requirement in terms of the network profile. Since the $face\_store$ doesn't
specify any requirement, it is deployed on the central cloud;

{\indent 3)} Only extracted faces (not the whole stream) are sent to the
cloud: this greatly reduces bandwidth consumption.

\noindent If another application using the same camera is added, graceful
degradation of performance occurs since available bandwidth is sufficient to
satisfy several concurrent applications.

\subsection{Privacy-aware Vehicles Tracking}
\label{sec:org7493b59}
In this scenario a tenant needs to track vehicles moving on certain streets
where cameras are installed, e.g., for the purpose of traffic flows
analysis. Tracking is performed by recognizing license plates \cite{openalpr}
and associating them with the location and the timestamp. Since this practice
involves privacy issues, licence plates are anonymized as soon as they are
collected.

Anonymization is performed by the \emph{anonymizer} microservice: data privacy
constraints suggest that the microservice should run on the local cloudlet
where the camera is installed. The anonymized data is then forwarded to a
second microservice, namely, the \emph{analyser}, hosted on the central cloud
and performing further analysis and final storage. This demonstration
comprises the following steps:

{\indent 1)} The tenant submits to Foggy two deployment requests:
$r_1=\{anonymizer, location\}$ and $r_2=\{analyzer\}$. Where $location$
indicates a geographical requirement specifying that the deployment has to be
done in a given region;

{\indent 2)} Foggy, accounting for the current status of the infrastructure,
deploys the microservices satisfying the tenant's requests. It deploys the
$anonymizer$ on the selected cloudlet in order to meet the location
requirement while the $analyzer$ is deployed on the cloud;

{\indent 3)} Only anonymized data are moved to the central cloud: this
guarantees data privacy;

{\indent 4)} In the case when a network fault happens between cloud and edge,
data is not lost because it is cached in the cloudlet and then, once
connectivity is restored, it is sent to the cloud (this would not be possible
without the presence of a cloudlet).

\section{Conclusion}
\label{sec:concl}
In this paper we have described Foggy, an architectural framework and a
software platform for workload orchestration and resource negotiation in a
multi-tier, highly distributed, heterogeneous and decentralized Cloud
Computing system. Through the presentation of three use cases, Foggy proves to
be able to orchestrate the workload in a Fog Computing environment. It acts as
a matchmaker between infrastructure owner and tenants improving i) the
efficient, effective and eventually optimized use of the infrastructure and
ii) the application performances to satisfy the requirements imposed.

After this initial deployment, we shall pursue, among the others, the
following aspects: i) a negotiation phase to involve also economic aspects
(i.e., pricing and billing) able to handle both tenants' and infrastructure
owners' needs; ii) modelling more complex interactions among different
application components; iii) different scheduling policies other than the FCFS
one; iv) new use cases to demonstrate real scenarios with multi-components
application in a way to measure impact and performance at infrastructure and
application level.


\bibliographystyle{IEEEtran} \bibliography{biblio}

\end{document}